\documentclass[useAMS,usenatbib,a4paper]{mn2e}
\usepackage[]{graphicx}

\def\approx{$\sim$}

\def\h2{$\rm H_2$}

\newcommand{\msun}{M$_{\odot}$}

\title[Comparing SFHs of the Magellanic Clouds from HST Imaging]{Comparing the Ancient Star Formation Histories of the Magellanic Clouds\thanks{Based on observations made with the NASA/ESA Hubble Space Telescope, obtained from the Data Archive at the Space Telescope Science Institute, which is operated by the Association of Universities for Research in Astronomy, Inc., under NASA contract NAS 5-26555}}
\author[Daniel R. Weisz]{Daniel R. Weisz$^{1}$ \thanks{E-mail: dweisz@astro.washington.edu} , Andrew E. Dolphin$^{2}$, Evan D. Skillman$^{3}$, Jon Holtzman$^{4}$, 
\newauthor
Julianne J. Dalcanton$^{1}$, Andrew A. Cole$^{5}$, Kyle Neary$^{3}$ \\
$^{1}$Department of Astronomy, Box 351580, University of Washington, Seattle, WA 98195\\ $^{2}$Raytheon, 1151 E. Hermans Road, Tucson, AZ 85756\\ $^{3}$Minnesota Institute for Astrophysics, University of Minnesota, 116 Church Street SE, Minneapolis, MN 55455, USA\\ $^{4}$Department of Astronomy, New Mexico State University, Box 30001, 1320 Frenger St., Las Cruces, NM 88003\\ $^{5}$School of Mathematics and Physics, University of Tasmania, Hobart, Tasmania, Australia}

\pagerange{\pageref{firstpage}--\pageref{lastpage}} \pubyear{2011}
\begin{document}

\maketitle
\label{firstpage}

\begin{abstract}

We present preliminary results from a new Hubble Space Telescope (HST) archival program aimed at tightly constraining the ancient ($>$ 4 Gyr ago) star formation histories (SFHs) of the field populations of the SMC and LMC.   We demonstrate the quality of the archival data by constructing HST/WFPC2-based color-magnitude diagrams (CMDs; $M_{F555W} \sim +$8) for 7 spatially diverse fields in the SMC and 8 fields in the LMC.  The HST-based CMDs are $>$ 2 magnitudes deeper than any from ground based observations, and are particularly superior in high surface brightness regions, e.g., the LMC bar, which contain a significant fraction of star formation and are crowding limited from ground based observations.  To minimize systematic uncertainties, we derive the SFH of each field using an identical maximum likelihood CMD fitting technique. We then compute an approximate mass weighted average SFH for each galaxy.  From the average SFHs, we find that both galaxies lack a dominant burst of early star formation, which suggests either a suppression or an under-fueling of ancient star formation in the MCs.  From 10-12 Gyr ago, we find that the LMC experienced a period of enhanced stellar mass growth relative to the SMC.   Similar to some previous studies, we find two notable peaks in the SFH of the SMC at $\sim$ 4.5 and 9 Gyr ago, which could be due to repeated close passages with the LMC, implying an interaction history that has persisted for at least 9 Gyr.  We find little evidence for strong periodic behavior in the lifetime SFHs of both MCs, suggesting that repeated encounters with the Milky Way are unlikely.  Beginning $\sim$ 3.5 Gyr ago, both galaxies show sharp increases in their SFHs, in agreement with previous studies.  Subsequently, the SFHs track each other remarkably well.  Spatial variations in the SFH of the SMC are consistent with a picture where gas was driven into the center of the SMC $\sim$ 3.5 Gyr ago, which simultaneously shut down SF in the outer regions while dramatically increasing the star formation rate in the center.  In contrast, the LMC shows little spatial variation in its ancient SFH.  The planned additional analysis of HST pointings at larger galacocentric radii will allow us to make more confident statements about spatial variations in the ancient SFHs of the SMC and LMC.

\end{abstract}


\newpage

\section{Introduction}
\label{intro}

The Magellanic Clouds (MCs) are among the best studied galaxies in the universe.  Their close proximity has motivated detailed observations of their gas, dust, and stellar contents \citep[e.g.,][]{zar97, kim98, zar02, sta04, kav06a, kav06b, mex06, uda08a, uda08b, ker09, gor11, mex11, rub12}, providing for a holistic understanding of their genesis and evolution.  In particular, stars fainter than the oldest main sequence turnoff (MSTO) are readily observable, providing for excellent constraints on the star formation histories (SFHs) of the MCs across all cosmic time.   In turn, such SFHs can be used as empirical discriminants between various evolutionary models of the MCs \citep[e.g.,][]{mur80, lin82, fic91, gar94, hel94, lin95, bek05, bes07, dia12, bes10, bes12}.

Despite extensive investments in measuring the SFHs of the MCs, our understanding of their ancient SFHs ($>$ 4 Gyr) is surprisingly limited.  Most wide-field ground based surveys provide spatially comprehensive coverage, but the resulting color-magnitudes diagrams (CMDs) only extend below the oldest MSTO in the uncrowded outer regions \citep[e.g.,][]{har04, har09, noe09, uda08a, uda08b, ker09, sah10, pia12, rub12}, leaving the ancient SFHs of more crowded inner regions highly uncertain.  Conversely, others have used the Hubble Space Telescope (HST) to overcome crowding limitations \citep[e.g.,][]{geh98, ols99, hol99, dol01, sme02, mcc05, cig12}, but such studies typically only consider small numbers of fields, compromising the spatially representative nature of the results.  Further, past SFH studies have typically focused on either the SMC or the LMC.  The use of different SFH measurement techniques or stellar libraries can introduce systematic offsets, making a comparison between the ancient SFHs of the SMC and LMC difficult.

\begin{figure}
\begin{center}

\includegraphics[scale=0.9]{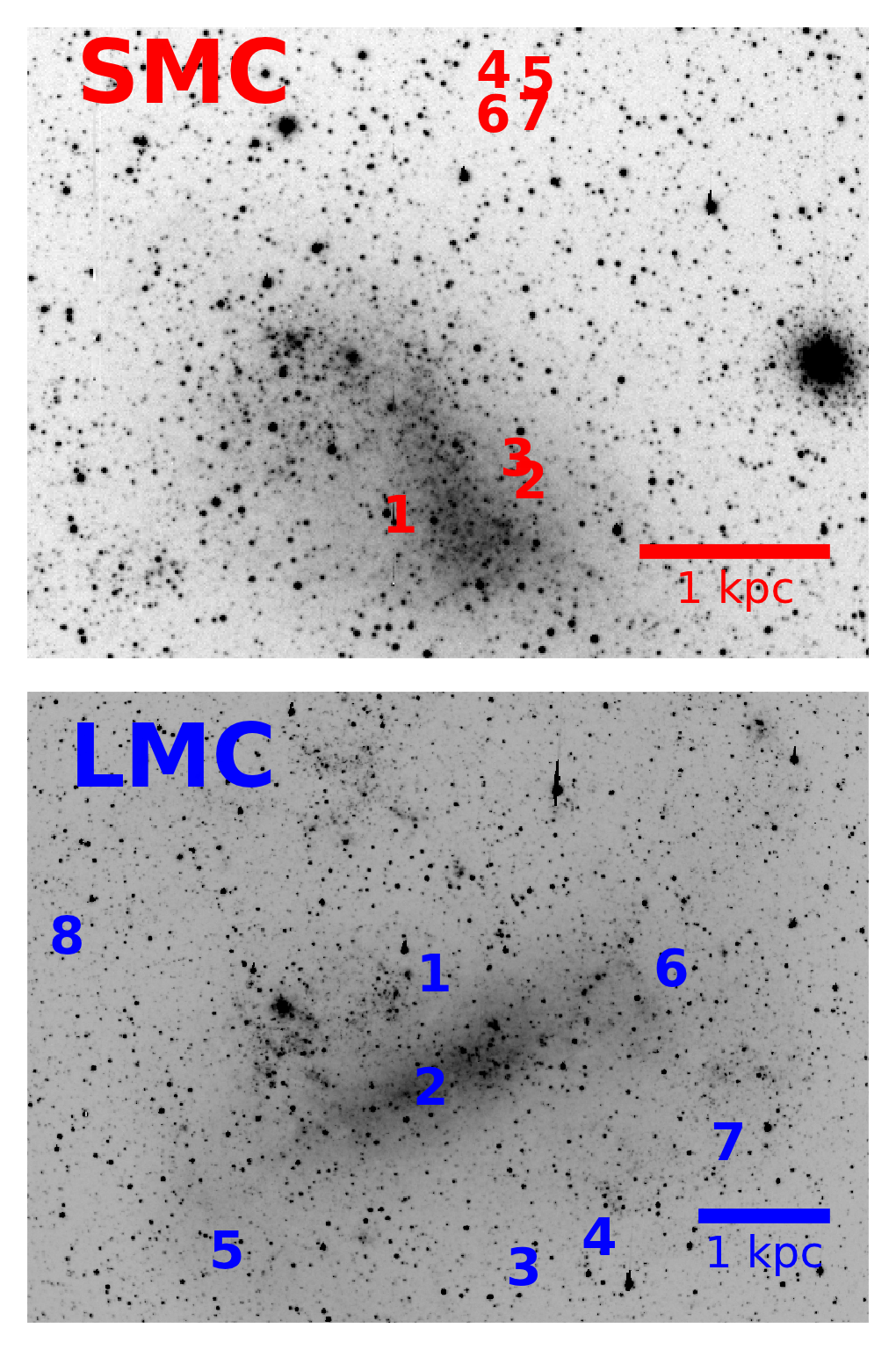}
\caption{The spatial locations of the HST/WFPC2 fields used in this study for the SMC (top) and LMC (bottom).}
\label{fig:mcs_spatial}
\end{center}
\end{figure}

To remedy these limitations, we have undertaken an HST/Wide Field Planetary Camera 2 \citep[WFPC2;][]{hol95} archival study aimed at self-consistently measuring the ancient SFHs of both MCs using over 100 HST fields (HST-AR-12853; PI. D.~Weisz).  Each of the CMDs selected for this study is significantly deeper than the oldest MSTO and reflect a diverse spatial sampling, enabling a precise and representative measurement of the ancient SFHs of the MCs.

In this paper, we present a first look at results from this ongoing archival program.   We have selected 7 spatially diverse HST fields in the SMC and 8 in the LMC to provide a preview of the data quality and to demonstrate the precision of the resulting SFH measurements.  We present the CMDs in \S \ref{sec:obs}, summarize the method of measuring SFHs in \S \ref{sec:measuresfh}, and present and discuss the derived SFHs in \S \ref{sec:sfhs}. The mapping between lookback time and redshift values used in this paper assume a standard WMAP-7 cosmology as detailed in \citet{jar11}. 

\begin{table*}
\begin{center}
\scriptsize
\begin{tabular}{cccccccccccc}
\hline 
\hline
  Field &  Field  &   RA &  DEC &   Galctocentric    &   \multicolumn{2}{c}{70\% Completeness}     &  No. Stars & HST-ID    &   Foreground & Differential & M$_{\star, total}$\\
Number & Name &  (J2000)  & (J2000)    & Radius (kpc)   & M$_{F555W}$    & M$_{F814W}$    & in CMD & & A$_{V}$ (mag)  & A$_{V}$ (mag) & (10$^{5}$~M$_{\odot}$)  \\
(1) & (2)  &  (3)  & (4)    & (5) & (6)  & (7)    & (8)   & (9) & (10) & (11) & (12)  \\
\hline 

       SMC-1 & u2o903 & 00:55:37 & $-$73:04:20 & 0.39 & $+$6.9 & $+$5.6 & 20315 & GO-6229 & 0.25 & $\ldots$ &  1.6 \\
       SMC-2 & u65c06 & 00:45:41 & $-$72:52:20 & 0.44 & $+$5.5 & $+$4.0 & 10563 & GO-8654 & 0.25 & $\ldots$ & 1.4\\
       SMC-3 & u46c01 & 00:46:40 & $-$72:44:43 & 0.52 &  $+$6.9 & $+$5.6 & 18294 & GO-6860 & 0.25 & $\ldots$ & 1.4 \\
       SMC-4 & u37704 & 00:45:54 & $-$70:34:43 & 2.3 &  $+$6.6& $+$4.8 & 818 & GO-6604 & $\ldots$ & $\ldots$ & $\ldots$   \\
       SMC-5 & u37706 & 00:48:54 & $-$70:47:43 & 2.3 &  $+$6.6 & $+$4.8 & 992 & GO-6604 & $\ldots$ & $\ldots$ & $\ldots$    \\
       SMC-6 & u377a4 & 00:46:06 & $-$70:46:44 & 2.3 &  $+$6.5 & $+$4.8 & 922 & GO-6604 & $\ldots$ & $\ldots$ & $\ldots$    \\
       SMC-7 & u377a6 & 00:48:54 & $-$70:32:44 & 2.3 & $+$6.6 & $+$4.8 & 783 & GO-6604 & $\ldots$ & $\ldots$ & $\ldots$   \\
       SMC-4-7  & 4-7 combined & $\ldots$ & $\ldots$ & $ 2.3 $ & $+$6.5 & $+$4.8 & 3515 & $\ldots$ & 0.15 & $\ldots$ & 0.3\\
       LMC-1 & u65006 & 05:24:06 & $-$68:48:48 & 0.1 &  $+$6.3 & $+$5.0 & 12971 & GO-8676 & 0.40 & 0.3 & 1.4\\
       LMC-2 & u2o901 & 05:24:17& $-$69:46:50& 0.86 &  $+$5.8 & $+$5.0& 30182 & GO-6229 & 0.25 & $\ldots$ & 3.7\\
       LMC-3 & u4b107 & 05:14:02 & $-$71:16:44 & 1.5 & $+$6.3 & $+$5.0 & 6149 & GO-7382 & 0.20 & $\ldots$ & 0.6 \\
       LMC-4 & u65005 & 05:06:20 & $-$70:58:21 & 1.6 &  $+$6.5 & $+$4.6 & 6400 & GO-8676 & 0.20 & 0.3 & 0.6\\
       LMC-5 & u65003 & 05:45:23 & $-$71:08:45 & 2.0 & $+$6.4 & $+$5.0 & 10258 & GO-8676 & 0.45 & 0.3 & 1.0\\
       LMC-6 & u63s01 & 05:01:56 & $-$68:37:20 & 2.0 &  $+$6.5 & $+$5.5 & 10781 & GO-8576 & 0.35 & 0.3 & 1.1\\
       LMC-7 & u65007 & 04:54:23 & $-$70:01:58 & 2.2 & $+$6.4 & $+$5.0 & 6358 & GO-8676 & 0.35 & 0.3 & 0.6\\
       LMC-8 & u2o902 & 05:58:18 & $-$68:20:51& 3.4 &  $+$7.3 & $+$5.5 & 3326 & GO-6229 & 0.25 & $\ldots$ & 0.2\\
\hline
\hline
\end{tabular}
\caption{\small The observational properties of the SMC and LMC fields.  The extinction values listed in columns 10 and 11 were derived from CMD fitting as described in \S \ref{sec:measuresfh}.   The total stellar mass formed in each field, i.e., the integral of the SFH, is listed in column 12.  As indicated, SMC fields 4-7 were combined to form a single larger field, SMC-4-7.}
\label{tab1}
\end{center}
\end{table*}

\section{The Data}
\label{sec:obs}

In this program, we utilize photometry and artificial star tests ($\sim$ 1.2$\times$10$^{5}$ per field)  taken from the Local Group Stellar Photometry Archive\footnote[1]{http://astronomy.nmsu.edu/holtz/archival/html/lg.html} \citep[LGSPA;][]{hol06}.  We focus on 7 fields in the SMC and 8 fields in the LMC, which were selected to represent the typical data quality and spatial distribution of the full archival program.  The locations of the fields are shown in Figure \ref{fig:mcs_spatial}.

\begin{figure*}
\begin{center}
\includegraphics[scale=0.8]{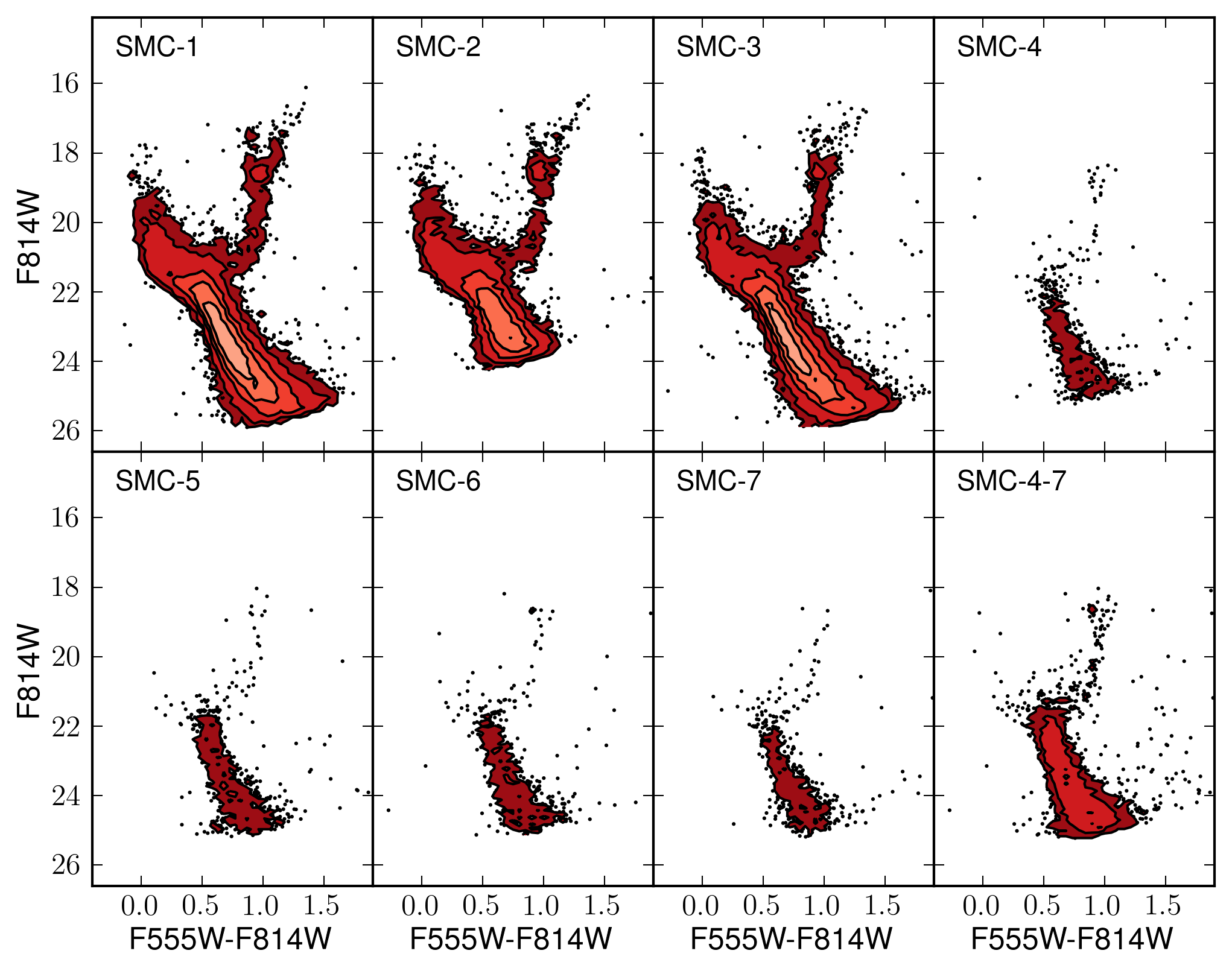}
\caption{HST/WFPC2-based CMDs of the 7 fields in the SMC.  The stellar density contours range from 2 (dark red) to 256 (light red) stars/decimag$^2$. The lower right hand panel represents the CMD for the combined outer SMC fields.  CMD characteristics such as number of stars and 70\% completeness limits are listed in Table \ref{tab1}.  These HST-based CMDs are typically $>$ 2 magnitudes deeper than those constructed from ground-based observations, particularly in high surface brightness regions where ground based observations are crowding limited.}
\label{fig:smc_cmds}
\end{center}
\end{figure*}

\begin{figure*}
\begin{center}
\includegraphics[scale=0.8]{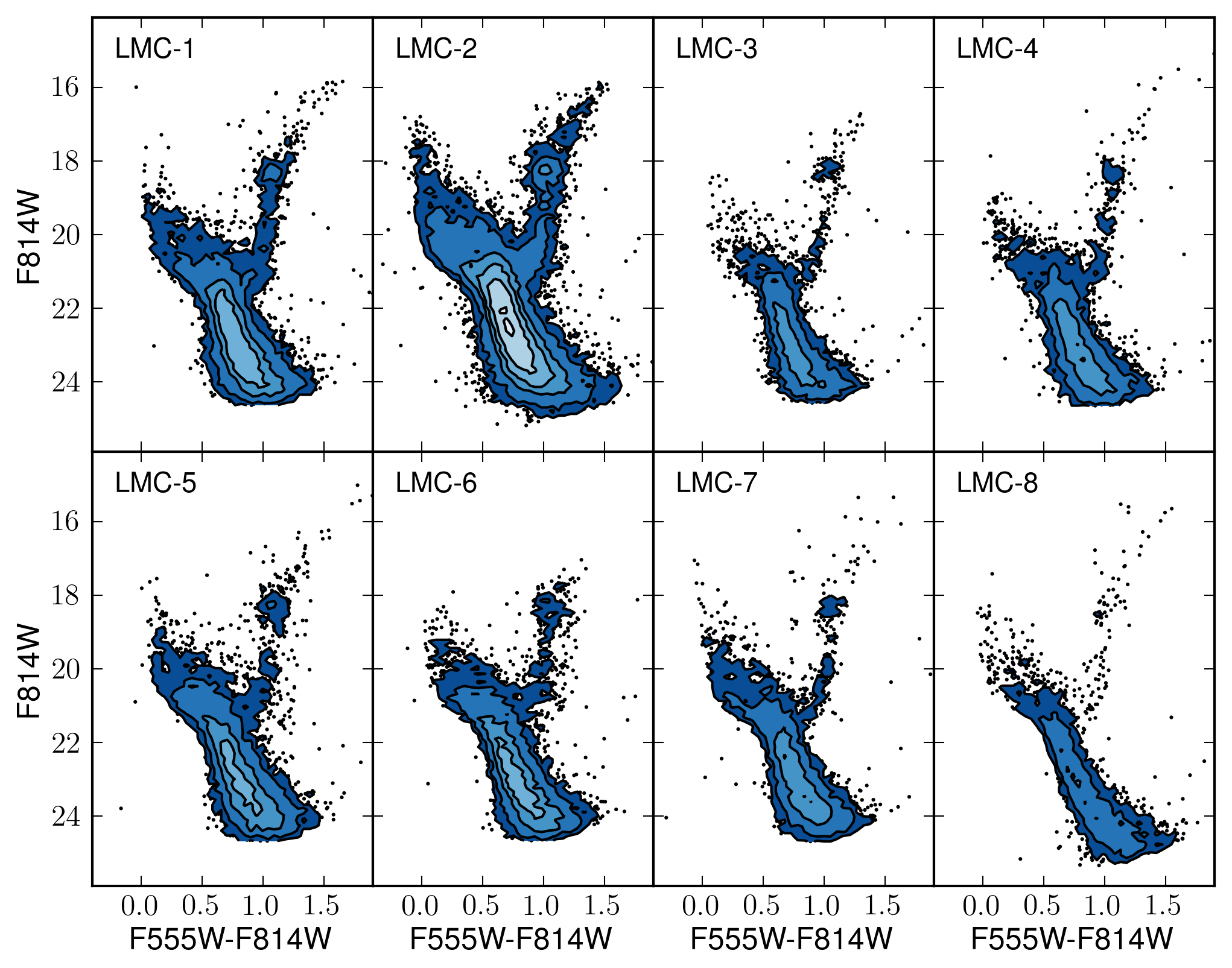}
\caption{HST/WFPC2-based CMDs of the 8 fields in the LMC.  The stellar density contours range from 2 (dark blue) to 512 (white) stars/decimag$^2$. CMD characteristics such as number of stars and 70\% completeness limits are listed in Table \ref{tab1}. These HST-based CMDs are typically $>$ 2 magnitudes deeper than those constructed from ground-based observations, particularly in high surface brightness regions, e.g., the LMC bar, where ground based observations are crowding limited}
\label{fig:lmc_cmds}
\end{center}
\end{figure*}

We have plotted the CMDs of the 7 SMC fields and 8 LMC fields in Figures \ref{fig:smc_cmds} and \ref{fig:lmc_cmds}, respectively.  These CMDs include only well-measured stellar sources, with flag values of 0 or 1 in the LGSPA database.   In general, each CMD only contains a few thousand stars, leaving some stellar sequences sparsely populated.  However, critical age sensitive features such as the oldest MSTO, sub-giant brach, red giant branch, and luminous main sequence are visually identifiable, providing confidence in the age leverage permitted by the data. To increase reliability of the SFHs, we merged the 4 sparsely populated outer SMC fields for subsequent analysis, after verifying the similarity of their completeness functions.

For both galaxies, HST-based CMDs are $>$ 2 magnitudes deeper than any CMDs from current ground based observations.  As discussed in \citet{hol06} and shown in Figures \ref{fig:smc_cmds} and \ref{fig:lmc_cmds} and in Table \ref{tab1}, all HST CMDs are highly complete down to their limiting magnitudes, which are in excess of $m_{F555W} =$ 24-26, even in the high surface brightness and crowded LMC bar (e.g., LMC-2).  In contrast, the current generation of ground based surveys have typical limiting magnitudes ranging from  $m_{V} \sim$ 22 in the disk to $m_{V} \sim$ 23-24 in the extreme outer halos \citep[e.g.,][]{zar97, har04, har09, noe09, ker09, uda08a, uda08b, sah10, rub12}.  Higher surface brightness regions, such as the LMC bar, tend to have substantially brighter magnitude limits, particularly when accounting for completeness effects \citep[e.g.,][]{har04, har09}, and are crowding limited from the ground. Overall, HST provides the best possible data for precise ancient SFH measurements in the MCs, particularly in high surface brightness regions,  modulo its limited spatial coverage relative to large ground based surveys.

\section{Measuring the Star Formation Histories}
\label{sec:measuresfh}

We measured the SFH of each field using MATCH, a maximum likelihood CMD fitting package \citep{dol02}.  Briefly, MATCH takes fixed input parameters (e.g., stellar evolution models, stellar initial mass function) and creates sets of synthetic simple stellar populations (SSPs).  These synthetic CMDs are convolved with observational errors from artificial star tests (ASTs) and combined to form a composite synthetic CMD.    Linear weights on the SSP CMDs are adjusted to obtain the most likely fit and correspond to the most likely SFH.  This process can be repeated at a variety of distance and extinction values to solve for these parameters as well.  A more detailed description of MATCH can be found in \citet{dol02}.

We quantified uncertainties in the SFHs using 50 Monte Carlo tests per field for a more detailed discussion of quantifying uncertainties in SFH measurements).  Given the comparably deep CMDs in both galaxies and the emphasis on relative SFHs of the two systems, we only computed the random uncertainties due to number of stars, and not the systematic uncertainties due to stellar evolution models.  We refer the reader to \citet{wei11} and \citet{dol12} for a more detailed discussion of error analysis in CMD-based SFH derivations.

All SFHs presented in this paper used the following parameters:  a Kroupa IMF \citep{kro01}, a binary star fraction of 0.35 with the mass of the secondary drawn from a uniform distribution, and the Padova stellar evolution models \citep{gir10}, with mass limits ranging from 0.15 to 120 \msun.  We adopted 40 logarithmic time bins between $\log(t)$ $=$ 6.6 to 10.15 with a single age bin for the very youngest stars, $\log(t)$ = 6.6-7.4, bins of 0.1 dex for  $\log(t)$ $=$ 7.4-9.0, and bins of 0.05 dex for $\log(t)$ $>$ 9.0.  We allowed the program to search for metallicities between $[M/H]$ $=$ $-$2.3 and 0.1 with a resolution of 0.1 dex.   Due to the exquisite depth of the CMDs, we have not placed any prior restrictions on the age-metallicity relationship in the SFH derivation process (e.g., we do not require a monotonically increasing metallicity toward the present).  Further, we allowed for a modest metallicity dispersion of 0.15 dex in each time bin, which helps account for potential metallicity spreads at a given age \citep[e.g.,][]{dac02}.  For SFHs derived from comparably deep CMDs, the resultant chemical evolution models are generally well-constrained and can coarsely discern chemical abundance variations as a function of time \citep[e.g.,][]{gal05, tol09}.

We initially allowed the program to solve for the distance to each field.  In each case we found the distances to be within $\pm$ 0.1 dex of commonly used distances in the literature.  For consistency in this preliminary study,  we therefore fixed the distance modulus for each field to 18.90 in the SMC and to 18.45 in the LMC \citep[][]{dol01, bon08}.  In the context of the larger archival dataset, we plan to further explore distance measurements from CMD fitting, which may provide orthogonal constraints to current 3D geometrical measurements of the MCs \citep[e.g.,][]{has12b, has12c}.

We limited our fits to only use stars brighter than the 70\% completeness limits as determined by $>$ 10$^{5}$ artificial star tests per field (see Table \ref{tab1}).  For LMC fields 3 and 6, we used slightly brighter limits, in order to exclude faint involved MS stars that are not included in the Padova models.  We also included a foreground CMD of model Milky Way stars following the CMD distributions derived in \citet{dej10}.

We allowed MATCH to solve for the line of sight extinction value for each field. The resulting extinction values were comparable with values from the extinction maps based on resolved star de-reddening \citep[][]{har97, zar02}.   Including modest amounts of differential extinction in the models improved fits of the elongated red clumps in several of the LMC CMDs (Fields 1, 4-7).  Specifically, we found a good differential extinction model to have the following characteristics: 50\% of the stars were modeled with $A_{V} =$ 0 and and 50\% with A$_{V}$ values evenly distributed between A$_{V}$ $=$ 0 and 0.3. 

\section{The Ancient Star Formation Histories of the Magellanic Clouds}
\label{sec:sfhs}

\subsection{Star Formation Histories of Individual Fields}
\label{sec:individualsfhs}

\begin{figure*}
\begin{center}
\includegraphics[scale=0.8]{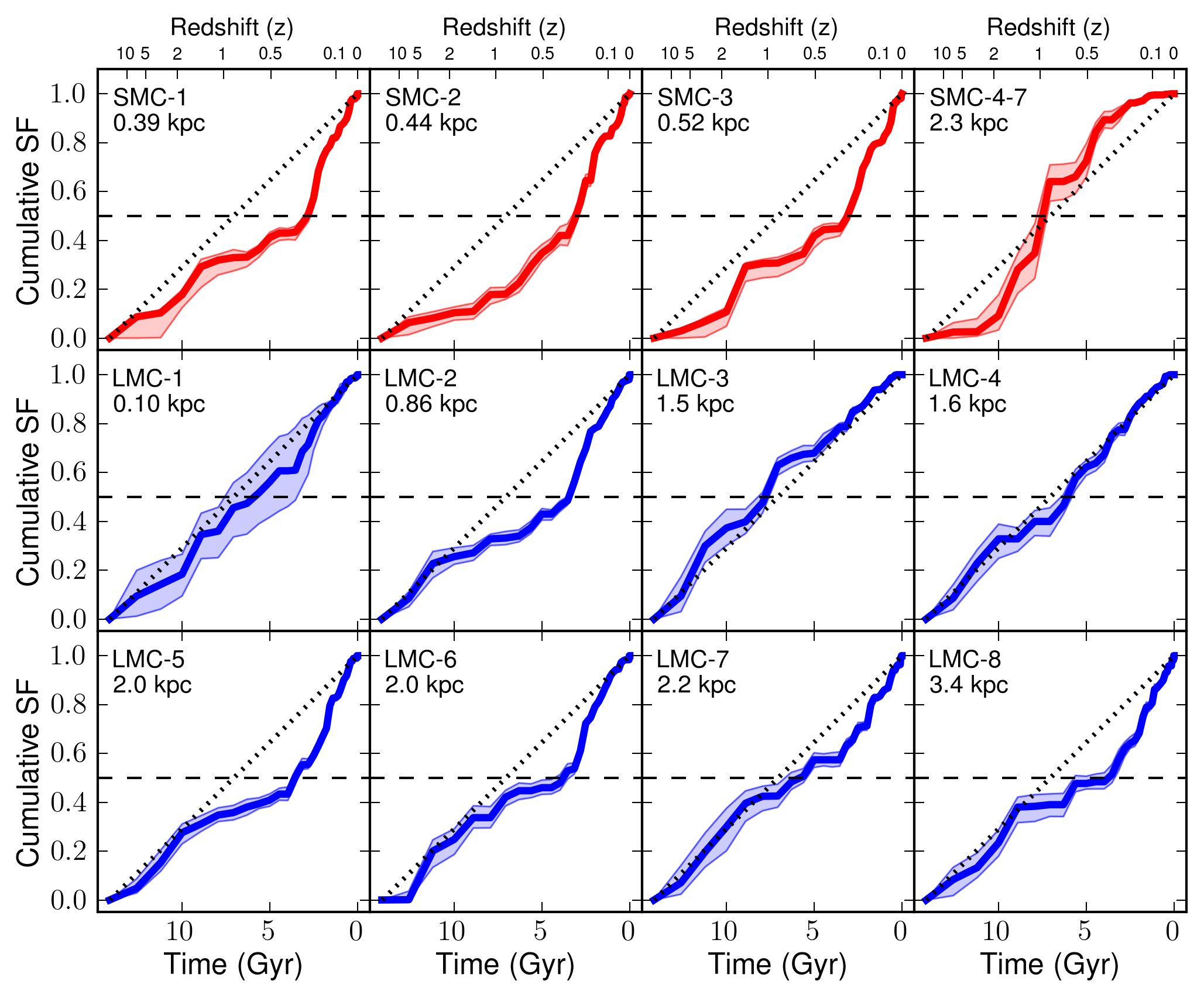}
\caption{The cumulative SFHs individual fields in the SMC (red) and LMC (blue).  The dark red solid line denotes the best fit SFH, while the lighter shaded envelope indicates the 1-$\sigma$ range of the random uncertainties.  The dot-dashed line represents the point at which 50\% of the total stellar mass formed, and the dotted line reflects a constant SFH.}
\label{fig:mc_sfhs}
\end{center}
\end{figure*}

In Figure \ref{fig:mc_sfhs}, we plot the normalized cumulative SFHs, i.e., the fraction of total stellar mass formed prior to a given epoch, for individual fields in the SMC and LMC.  For the SMC, we see that each field formed less than 10\% of its total stellar mass prior to $\sim$ 12 Gyr ago.  The outer fields (4-7) of the SMC formed the bulk of their stars between $\sim$ 5 and 8 Gyr ago, with little subsequent star formation (SF) until the present, a similar finding to outer area SMC SFHs derived by \citet{dol01} and \citet{noe09}.   In contrast, the inner fields show a nearly constant  SFH from $\sim$ 3.5-12 Gyr ago, followed by a dramatic increase in SF $\sim$ 3.5 Gyr ago, just when SF in the outer galaxy has shut down.  Differences in the SFHs of the inner and outer fields are consistent with previously observed population gradients in the SMC \citep[e.g.,][]{noe09}.  

Like the SMC, most fields in the LMC formed a small percentage of their total mass prior to $\sim$ 12 Gyr ago.  Subsequently, most fields experienced a nearly constant SFH until $\sim$ 4 Gyr.  At this point, approximately half the fields show signs of an increase in their SFHs, while the other half continued forming stars at a nearly constant rate.  There is a slight spatial correlation such that fields close to the bar appear to preferentially show the a rise in star formation, while those farther away do not.  However, the majority of the individual field SFHs are consistent at the 1-$\sigma$ level, suggesting a relatively weak population gradient in the LMC over the radial extent subtended by our HST fields, and with previous HST-based studies \citep[e.g.,][]{geh98}.  Our derived SFHs for Fields 2 and 8 are qualitatively similar to previous analysis of the same data \citep{hol99}.

\subsection{The Mean Star Formation Histories}
\label{sec:meansfhs}

To compare SFHs between the LMC and SMC, we compute the weighted mean SFH for each galaxy, using the total mass formed in each field (listed in Table \ref{tab1}) as a weight.  This approach ensures that the mean SFH is a reasonable proxy for the SFH of the entire galaxy, i.e., integrating over the mean SFH provides a proxy for the total stellar mass of each galaxy, modulo the inherent uncertainty in extrapolating the mean from a small set of fields.  More formally, the mean mass weighted SFH from a set of individual fields can be written as

\begin{equation}
\overline{x} = \frac{\sum_i^N w_i \, \mathrm{x}_i}{\sum_{i}^{N} w_i} \, ,
\label{eq:weightedmean}
\end{equation}

\noindent where $x_i$ is the SFH of a single field and $w_i$ is the total stellar mass formed in that field.  Similarly, the 1-$\sigma$ uncertainties on the weighted mean SFH are defined to be

\begin{equation}
\sigma(\overline{x}) = \sqrt{\frac{\sum_i^N w_i^2 \, \sigma_i^2}{\sum_i^N w_i^2} }\, ,
\label{eq:weightedsigma}
\end{equation}

\noindent where $\sigma_i$ is the set of uncertainties on the SFH of an individual field.  This formalism assumes that the error bars on the SFHs are independent, which is valid in this context as we only consider random uncertainties (which are independent) and not systematic uncertainties (i.e., due to stellar evolution model selections, which are not independent).  We refer the reader to Appendix C of \citet{wei11} and to \citet{dol12} for further discussion of random and systematic uncertainties in SFHs and their correct propagation for ensemble average SFHs.  In addition to propagating uncertainties, we also computed the dispersion in the best fit SFH of each field and added the result in quadrature to the errors in Equation \ref{eq:weightedsigma}.

\begin{figure}
\begin{center}
\includegraphics[scale=0.8]{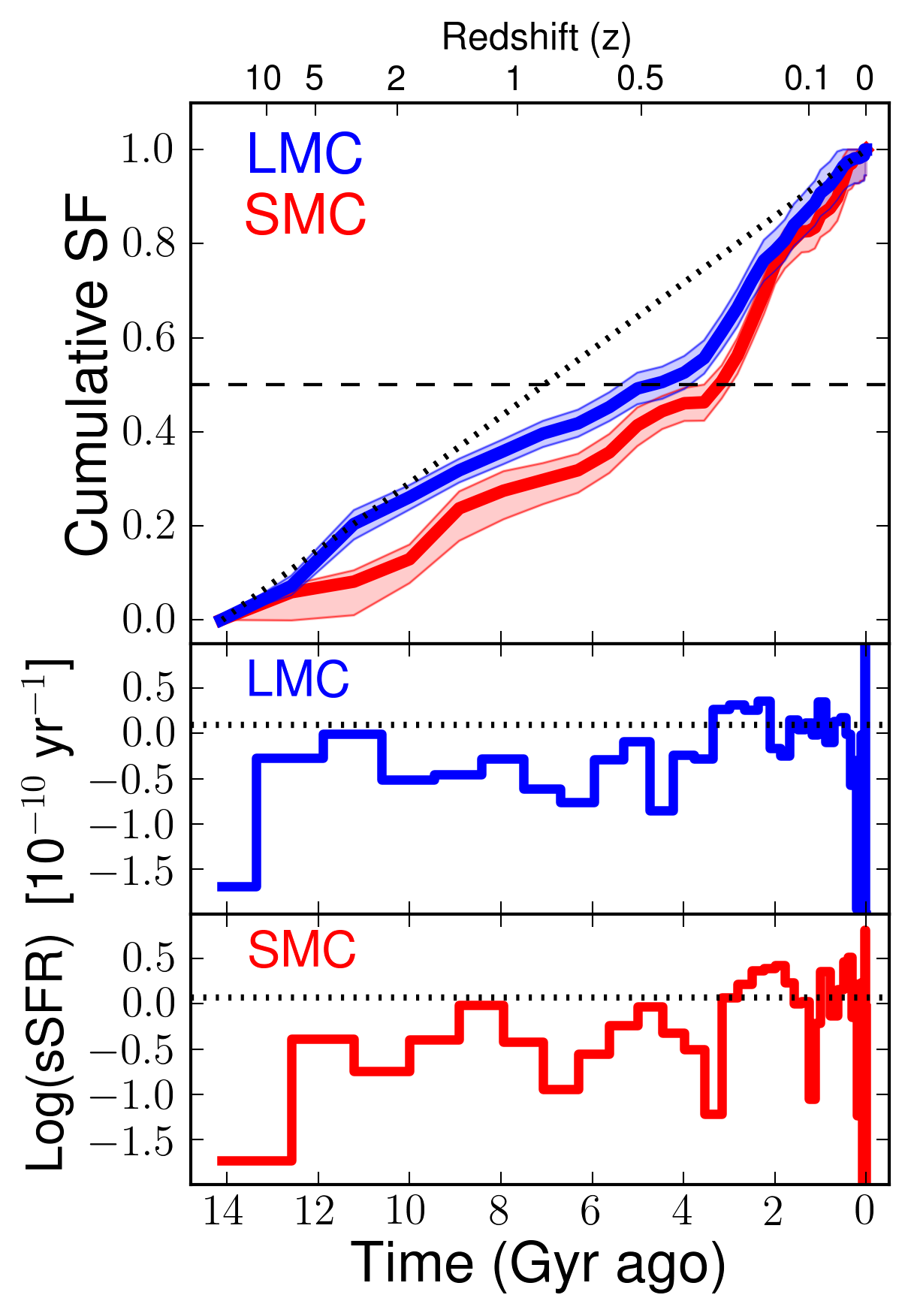}
\caption{The sample averaged cumulative SFHs for the SMC (red) and LMC (blue) in the top panel and, for increased clarity, the mean specific SFHs in the lower panels.  The shaded error envelopes represent the standard error in the mean cumulative SFHs due to random uncertainties and the dispersion in the best fits from the individual SFHs.  The black dotted line reflects a constant lifetime SFH, and the dot-dashed line represents the formation of 50\% of the total stellar mass.  From $\sim$ 3.5-12 Gyr ago, both the SMC and LMC show constant SFHs at levels lower than the lifetime averages.  The SFH of the SMC has two peaks at $\sim$ 4.5 and 9 Gyr ago.  Both SFHs show a dramatic increase starting around $\sim$ 3.5 Gyr ago, a feature consistent with previously derived SFHs of the MCs, and are subsequently remarkably similar.  The globally representative nature of these SFHs for  ages $<$ 2 Gyr are uncertain due to the mixing timescale of young populations and the limited number of HST fields used in this paper.}
\label{mcsfhs}
\end{center}
\end{figure}

As shown in Figure \ref{mcsfhs}, the resulting average SFHs of the MCs exhibit several interesting features.  First, for ages $>$ 12 Gyr ago, both systems having formed identical fractions ($\sim$ 10\%) of their total stellar mass.  Second, from $\sim$ 10-12 Gyr ago, the LMC shows an enhanced period of mass growth relative to the SMC.  Third, from $\sim$ 3.5-12 Gyr ago, both galaxies show mostly constant SFHs, with occasional factor of a few increases compared to the average SFR during this interval (e.g., at $\sim$ 11.5 Gyr ago for the LMC and at $\sim$ 4.5 and 9 Gyr ago for the SMC). Fourth, we find that the SMC and LMC both show sharp increases in their typical SFRs starting $\sim$ 3.5 Gyr ago to the present, such that they form 45-55\% of their mass at these recent times.  Finally, the SFHs both galaxies track each other very well from 3.5 Gyr ago to the present.

\subsection{Comparison to Previously Derived Star Formation Histories}
\label{sec:meansfhs}

Our mass weights mean SFHs are similar to SFHs derived in previous studies.  In the case of the LMC, previous HST-based SFHs found that the LMC formed 50\% of its mass $\sim$ 5-6 Gyr ago and shows a dramatic rise in SFR $\sim$ 3.5-4 Gyr ago \citep[e.g.,][]{geh98, ols99, hol99, sme02}, features also found by \citet{har09} using shallower data.  For the SMC, previous HST-based studies found that the SMC formed $\sim$ 50\% of it stellar mass around $\sim$ 3-4 Gyr ago and experienced a dramatic increase in SF $\sim$ 3.5-4 Gyr ago \citep[e.g.,][]{noe09, cig12}, both similar to results from \citet{har04}.  The SFHs of the outer regions ($R$ $>$ 3 kpc) of the SMC also show a similar increase in SF $\sim$ 8-9 Gyr ago \citep[e.g.,][]{dol01, noe09} to our SFH.  Completed analysis of the full compliment of archival fields will enable more precise global and spatially resolved comparisons to other MC SFH studies.

\subsection{Implications for Evolutionary Scenarios of the Magellanic Clouds}
\label{sec:evolution}

Well-constrained SFHs can be useful in discriminating between MCs evolutionary scenarios.  Past interactions (or lack thereof) with one another and/or the Milky Way may have affected the intensity and duration of past star formation events, which can be extracted from precisely constrained lifetime SFHs of both galaxies.  In this section, we briefly discuss how our derived SFHs can be used in conjunction with other observations to help discern between various proposed models for the genesis and evolution of the MCs.

At the oldest times ($>$ 12 Gyr ago), the MCs exhibit a level of star formation consistent with a constant lifetime SFH.  Such constant star formation at early times makes the MCs standouts among satellites of comparable distance to the Milky Way (e.g., Ursa~Minor, Draco), which have predominantly old and truncated SFHs \citep[e.g.,][]{dol05, tol09}.  Instead, the constant level of ancient star formation in the MCs more closely resemble isolated dwarfs in the Local Group \citep[e.g., Leo~A, IC~1613; ][]{col07, ski03}, each of which have deep CMDs enabling similar SFH derivations (i.e., no age-metallicity restriction was necessary).  This finding qualitatively suggests that the MCs were located in a more isolated or possibly field environment at early stages in their evolution.  Alternatively, the low levels of star formation may be due to a lack of sufficient fuel at early times.  For example, the metallicites of RR Lyrae in the LMC are consistent with an early and rapid chemical enrichment scenario \citep[e.g.,][]{has12a}, which is better matched to an instantaneous gas recycling model as opposed to significant accretion of lower metallicity material \citep[e.g.,][]{pag98, car08a}.  Studies by \citet{pia05} and \citet{car08b} suggest a similar rapid early enrichment scenario is also possible in the SMC.

From $\sim$ 3.5 to 12 Gyr ago, there are interesting features in the SFHs of the MCs.  First, from $\sim$ 10-12 Gyr ago, there is a pronounced difference in the SFHs of the SMC and LMC.  Over this interval, the LMC experienced a substantial period of mass growth, forming a significantly larger fraction of its total stellar mass compared to the SMC.  This suggests that the LMC may have  accreted more material and/or formed stars more efficiently than the SMC at this time.  Subsequent to this interval, the SFHs of the MCs do not exhibit such drastic differences.

Second, from $\sim$ 3.5 and 10 Gyr ago, there are subtle differences between the SFHs of the two galaxies.  Over this period, the SFH of the LMC was nearly constant, while the SFH of SMC appears to have peaks at $\sim$ 4.5 and 9 Gyr ago.  These star formation enhancements in the SMC could be be the result of interactions with the LMC.  Due to the order of magnitude mass difference in the systems, an interaction between the two systems would like enhance the global SFR of the SMC, but would not similarly affect the LMC \citep[e.g.,][]{con04, col05, bes12}. 

These peaks may have implications for the dynamical history of the two systems.  In the simplest two body interpretation, the SMC's SFH suggests that the SMC and LMC have been bound for at least the past $\sim$ 9 Gyr, i.e. the age of the first peak, with an orbial period of $\sim$ 4.5 Gyr, i.e, the time of the second peak.  However, simulations suggest that is is unlikely for a MC-like pair to be stable for more than $\sim$ 5 Gyr \citep[e.g.,][]{bek05, dia12}.  Further, observational studies find that SMC-LMC-Milky Way triplets are exceedingly rare in the Local Volume \citep[e.g.,][]{liu11, rob12}, although LMC-Milky Way pairs appear slightly more common \citep[e.g.,][]{tol11}.  Taken at face value, it appears that the enhancements in the intermediate age SFH of the SMC are either unrelated to the long term dynamical history of the MCs or that the MCs are extreme outliers in the timescale distribution of stable binary systems.  In the case of the former, the origin of the seemingly periodic star formation enhancements in the SMC is unclear, while for the latter, a long lived SMC-LMC binary is somewhat easier to explain if the pair had a late infall time.  

A third clear SFH feature is the increase in star formation in both galaxies $\sim$ 3.5 Gyr ago, presumably due to the gravitational influence of the Milky Way \citep[e.g.,][]{mur80, gar94, bek05}.  These strong increases in star formation activity are unprecedented in either galaxy, suggesting that this is episode is likely due to a first close passage with the Milky Way, as opposed to a history of repeated orbits.   Subsequent interactions between the two MCs may have triggered smaller peaks in the SFH of the SMC and/or transferred stars and gas between the systems \citep[e.g.,][]{ols11, bes12}.

Finally, we connect the variations in the SFHs of individual fields to putative historical interaction scenarios of the galaxy pair.  The presence of age gradients in the SMC \citep[e.g.,][]{gar92, zar00, sab09} and LMC \citep[e.g.,][]{har09, sah10, pia12} are well-documented.  In the case of the SMC, our SFHs show similar behavior to previous studies.  Namely, at $\sim$ 3.5 Gyr ago, star formation in the SMC has shut off in the outer regions and sharply increased in the inner regions.  One interpretation is that the gas supply of the lower mass SMC was funneled into the center of the galaxy, presumably due to an interaction with the more massive LMC.  Models suggest that the MCs were $\sim$ 500 kpc from the Milky Way 3.5 Gyr ago \citep[e.g.,][]{bes07}, making other physical effects such as ram pressure stripping less likely.  Alternatively, our outer SMC fields may in fact simply be probing the older `halo' population of the SMC \citep[e.g.,][]{nid11}.  Overall, the uncertainties in the geometry of the SMC, i.e., edge-on vs. face-on, and the small number of fields considered in this preliminary study make it challenging to causally interpret spatial SFH variations in the SMC.

In both galaxies, our fields do not extend to the `halo' fields.  At such large galactocentric distances, the stellar crowding is typically minimal and deep CMDs can be constructed from ground based imaging \citep[e.g.,][]{noe09, sah10}.  As a results of our disk biased spatial coverage the population gradient we find in the SMC and lack of population gradient in the LMC are not globally representative.  Completed analysis of the full complement of archival fields will significantly expand the extent of the HST spatial coverage, relative to this study, and allow us to explore the connection between global and spatially-varying SFHs scenarios in greater detail.

\section{Summary}
\label{summary}

We have presented preliminary results from an HST archival program aimed at constraining the ancient field population SFHs of the MCs (HST-AR-12853; PI. D.~Weisz).  To demonstrate the quality of the data and preview results of the full program, we have plotted the CMDs 7 spatially diverse fields in the SMC and 8 fields in the LMC and used an identical CMD fitting technique to derive the SFHs of each field.  From the SFHs of the individual fields, we computed the mass weighted average SFH for each galaxy, and found them to be in good agreement with those from previous studies.  From the average SFHs we found that (1) for ages $>$ 12 Gyr, both galaxies exhibit star formation consistent with a constant lifetime SFH, suggesting either suppressed or under-fueled star formation, relative to other Milky Way satellites at comparable distances; (2) the LMC shows enhanced mass growth from $\sim$ 10-12 Gyr ago relative to the SMC; (3) the SMC shows distinct peaks in its SFH $\sim$ 4.5 and 9 Gyr, which may be due to periodic encounters with the LMC, while the LMC had roughly constant star formation at the same epochs.  Assuming this is an interaction drive feature, it implies an interaction history that has persisted for at least 9 Gyr; (4) at $\sim$ 3.5 Gyr ago both galaxies show a sharp increase in SFR, and their SFHs track each other subsequently, consistent with suggestions of a close encounter with the Milky Way at recent times; (5) Starting around $\sim$ 3.5 Gyr ago, the SF in the outer regions of the SMC abruptly ceases, while the SFR in the galaxy's center sharply increases, suggesting gas from the outer regions has been centrally funneled.  Interpretation of this finding is somewhat complicated to uncertainties in the geometry of the SMC.  In contrast, the LMC shows only small spatial variations in its SFH over the radial extent covered by our HST fields.  In both cases, our interpretation of spatial trends is limited by the disk-biased spatial coverage of the HST fields considered in this paper.  Our planned analysis of $>$ 100 HST archival fields will increase the radial coverage, enabling more secure statements about spatial variations in the SFHs of the MCs and more detailed comparisons with various MC evolutionary models.

\section*{Acknowledgements}

The authors would like to thank the anonymous referee for insightful comments that helped improve the scope and clarity of this paper.  The authors would also like to thank Gurtina Besla for her close reading of the manuscript and helpful comments on the current state of Magellanic Cloud evolutionary models.  Support for this work was provided by NASA through grant number HST GO-12853 from the Space Telescope Science Institute, which is operated by AURA, Inc., under NASA contract NAS5-26555. This research made extensive use of NASA's Astrophysics Data System Bibliographic Services.

\bsp

\end{document}